\documentclass[aps,prb,twocolumn,superscriptaddress,showpacs]{revtex4}

\usepackage{amsbsy}
\usepackage{bm}
\usepackage{graphicx}
\usepackage{amssymb}

\newcommand{\smeq}{\! = \!}
\newcommand{\smneq}{\! \neq \!}
\newcommand{\smpl}{\! + \!}
\newcommand{\smmi}{\! - \!}

\newcommand{\be}{\begin{equation}}
\newcommand{\ee}{\end{equation}}
\newcommand{\bea}{\begin{eqnarray}}
\newcommand{\eea}{\end{eqnarray}}

\newcommand{\ci}{\mathrm{i}}

\begin{document}

\title{Magnetic breakdown of cyclotron orbits in systems with Rashba and Dresselhaus spin-orbit coupling}

\author{A. A. Reynoso}

\affiliation{Instituto Balseiro and Centro At\'{o}mico Bariloche,
Comisi\'{o}n Nacional de Energ\'{\i}a At\'{o}mica, 8400 S. C. de
Bariloche, Argentina.} \affiliation{Consejo Nacional de
Investigaciones Cient\'{\i}ficas y T\'ecnicas (CONICET),
Argentina}

\author{Gonzalo Usaj}
\affiliation{Instituto Balseiro and Centro At\'{o}mico Bariloche,
Comisi\'{o}n Nacional de Energ\'{\i}a At\'{o}mica, 8400 S. C. de
Bariloche, Argentina.} \affiliation{Consejo Nacional de
Investigaciones Cient\'{\i}ficas y T\'ecnicas (CONICET),
Argentina}

\author{C. A. Balseiro}
\affiliation{Instituto Balseiro and Centro At\'{o}mico Bariloche,
Comisi\'{o}n Nacional de Energ\'{\i}a At\'{o}mica, 8400 S. C. de
Bariloche, Argentina.} \affiliation{Consejo Nacional de
Investigaciones Cient\'{\i}ficas y T\'ecnicas (CONICET),
Argentina}

\begin{abstract}

We study the effect of the interplay between the Rashba and the Dresselhaus spin-orbit couplings on the transverse electron focusing in two-dimensional electron gases. Depending on their relative magnitude, the presence of both couplings can result in the splitting of the first focusing peak into two or three. This splitting has information about the relative value of spin-orbit couplings and therefore about the shape of the Fermi surface. More interesting, the presence of the third peak is directly  related to the tunneling probability ("magnetic breakdown") between orbits corresponding to the different sheets of the Fermi surface. In addition, destructive interference effects between paths that involve tunneling and those that do not can be observed in the second focusing condition. Such electron paths (orbits) could be experimentally detected using current techniques for imaging the electron flow opening the possibility to directly observe and characterize the magnetic breakdown effect in this system.

\end{abstract}
\date{\today}

\maketitle

\section{Introduction}
Transport properties of two-dimensional electron (2DEG) and hole
gases can be affected in very peculiar ways by the spin-orbit (SO)
coupling. The unusual properties of spin transport of these
systems are seen as promising tools for the development of new
spintronic devices,\cite{Spintronicsbook} which would allow us to
coherently control and manipulate the electrons' spin. This
triggered an intense activity in the field during the past years.
Among the SO-related effects in 2DEGs, it is worth mentioning the
proposal of a spin filtering
transistor,\cite{DattaD90,SchliemannEL03} the Aharonov-Casher
oscillation in mesoscopic rings,\cite{KonigTHSHDSBBM06,KovalevBJMS07} and the spin Hall
effect.\cite{DyakonovP71,Hirsch99,MurakamiNZ03,SinovaCSJM04,KatoMGA04,UsajB05_SHE,WunderlichKSJ05,KatoMGA05-lshaped,
SihMKLGA05,NikolicSZS05,NomuraWSKMJ05,EngelHR05,ErlingssonL05,NomuraSJNM05,AdagideliB05,ReynosoUB06}

In 2DEGs made either from heterostructures or from quantum wells, there
are two dominating forms of the SO coupling.\cite{Winkler_book}
The Rashba SO coupling, which arises from the asymmetry of the
confinement potential of the 2DEG, and the Dresselhaus SO
coupling, which arises from the lack of inversion symmetry of the
crystal structure. Both types of SO couplings are present in
general and their relative magnitude depends on the structure and
the materials used to make the 2DEG. The SO couplings are
relatively weak in AlGaAs-GaAs structures and quite strong in In
or Sb based semiconductors. There are two important differences
between the Rashba and Dresselhaus SO couplings. On the one hand,
the magnitude of the former can be externally controlled by a gate
voltage,\cite{NittaATE97,KonigTHSHDSBBM06}
providing a interesting new knob to control transport properties.
On the other hand, the Rashba coupling is isotropic while the
Dresselhaus coupling depends on the orientation of the crystal
axes. These two different sources of the SO coupling can be experimentally determined using different techniques.
\cite{Ganichev2004,Giglberger2007,MillerZMLGCG07,MeierSSGSE07,MeierSGSE08}

Both (linear) SO couplings lead to similar electronic and
transport properties when one dominates. However, very
interesting effects arise when the two couplings have a similar
magnitude. In particular, when both couplings are equal the spin
and the momentum decouple. This effect has been proposed as a way
to built up a spin transistor in disordered systems.
\cite{SchliemannEL03} In addition, it was argued that in that
case the system contains unusually long-lived spin excitations.\cite{BernevigOZ06} Indications of the presence of this excitations have
been observed very recently.\cite{WeberOBZSA07} Also, the magnetic
field anisotropy of the spin relaxation length in long wires made
from 2DEGs in AlGaAs has been attributed to the closed values of
the two couplings.\cite{FrolovVYLWF08}

It is therefore interesting to look for new alternatives where the
effect of the competition between the Rashba and Dresselhaus
couplings on the transport properties can be measured directly. In
this work, we analyze the effect of such competition on the
transverse electron focusing
signal.\cite{vanHouten89,BeenakkerH91} In Ref.
[\onlinecite{UsajB04_focusing}] it was predicted that SO coupling leads to
the splitting of the odd focusing peaks. Since then, this
splitting has been observed in different samples
\cite{RokhinsonLGPW04,DedigamaDMGKSSMH06} and discussed
by several authors. \cite{ReynosoUSB04,ZulickeBW07,Schliemann2008}
Here, we show that the splitting of the focusing peaks can be used to map out the non-trivial shape of the Fermi surface of the 2DEGs when both
types of linear SO couplings are present and have a similar
magnitude. In addition, we found that the focusing experiment can
clearly show the tunneling between cyclotron orbits, in direct
analogy to the magnetic breakdown in bulk
materials.\cite{CohenF1961,AndradaSilvaLRB94}

To our knowledge, this is the first example where the magnetic
breakdown between different cyclotron orbits could be directly observed. This could be done by  using, for instance, the imaging technique developed by Westervelt and
coworkers.\cite{TopinkaLSHWMG00,TopinkaLWSFHMG01,AidalaPHW06,Reynoso2006b,AidalaPKHWHG07}

\section{Spin-orbit coupling in two dimensional gases}
\subsection{Bulk eigenstates}
The Hamiltonian of a 2DEG in the presence of both Rashba and
Dresselhaus SO couplings is given by 
\be
H=\frac{
p^2}{2m^*}+\frac{\alpha}{\hbar}(p_y\sigma_x-p_x\sigma_y)
+\frac{\beta}{\hbar}(p_{x'}\sigma_{x'}-p_{y'}\sigma_{y'})
\label{Hfree}
\ee
where $\bm{p}=(p_x,p_y)$ is the momentum
operator, $\alpha$ and $\beta$ are the Rashba and Dresselhaus
coupling parameters, respectively, and $\{\sigma_i\}$ are the Pauli
matrices. The axes $x'$ and $y'$ correspond to crystallographic
directions while $x$ and $y$ are arbitrary directions chosen in a
convenient way---note that the Rashba term is isotropic and
therefore independent of the axes choice.

\begin{figure}[t]
 \centering
\includegraphics[width=0.28 \textwidth,clip]{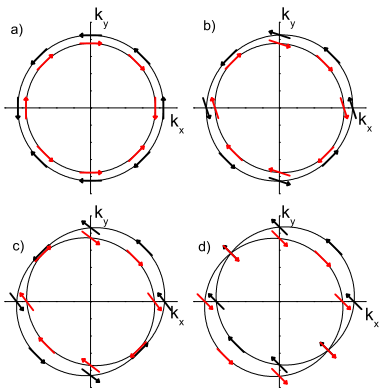}
  \caption{The Fermi surfaces for $\beta/\alpha\smeq0,0.5,0.75$ and $1$ are shown in (a), (b), (c) and (d), respectively. The arrows indicate the spin orientation of the corresponding eigenstate.}
 \label{FermiSurface}
\end{figure}

Hamiltonian (\ref{Hfree}) can be easily diagonalized by
proposing a solution in the form of a plane wave. The eigenfunctions and eigenvalues are
\bea
\Psi_{\bm{k}}^\pm(\bm{r})&=&\frac{1}{\sqrt{2 A}}\mathrm{e}^{\ci\bm{k}\cdot\bm{r}}\left( {{\pm 1 \atop {\rm e}^{{\ci}\phi }}}\right)\\
\varepsilon(\bm{k})&=&\frac{\hbar^2 k^2}{2m^*}\!\pm\!\sqrt{(\alpha
k_x\smpl\beta k_y)^2\smpl(\alpha k_y\smpl\beta k_x)^2}\,.
\label{band}
\eea
Here,
$\tan\phi=-(\alpha k_x+\beta k_y)/(\alpha k_y+\beta k_x)$, $A$ is
the system's area and we choose $x=x'$ and $y=y'$. Figure
\ref{FermiSurface} shows the corresponding Fermi surface for
different values of the ratio $\beta/\alpha$. The arrows indicate
the spin orientations of the eigenstates.

The competition between the Rashba and Dresselhaus SOs originates
a deviation of the Fermi surface from the circular shape. For
$\alpha=\pm\beta$ the Fermi surfaces recover a circular shape
shifted from the $\Gamma$ point (there is a perfect nesting
between the two surfaces) and the spin orientation becomes
independent of $\bm{k}$---nevertheless, this case has interesting
spin properties.\cite{SchliemannL03,MishchenkoH03,SchliemannEL03,BernevigOZ06}
For $|\beta/\alpha|\smneq1$ the two different Fermi surfaces do not cross each other.
The minimum and maximum distances between them in $k$-space are given by $\Delta k_\pm=\frac{2m^*}{\hbar^2}|\alpha\pm\beta|$. As we will show, these properties have important consequences on the transverse focusing signal .

When an external magnetic field is applied perpendicular to the sample, Landau levels are formed.
In that case, a Zeeman term must be included in Hamiltonian
(\ref{Hfree}) and $\bm{p}$ should be replaced by $
\bm{p}+(e/c)\bm{A}$, with $\bm{A}$ as the vector potential. A closed
analytical solution for arbitrary values of $\alpha$ and $\beta$
is not known (see however Refs. [\onlinecite{ZareaU05,Zhang06}]).
For $\beta=0$, however, a straightforward
calculation\cite{BychkovR84} shows that the energy spectrum is
given by $E_{n}^{\pm }\!=\!\hbar \omega _{c}n\mp
(E_{0}^{2}+(\alpha/l_{c}) ^{2}2n)^{\frac{1}{2}}$, where $n\geq 1$,
$\omega _{c}\!=\!e\left| B\right| /m^{*}c$ is the cyclotron
frequency, $l_{c}\!=\!(\hbar /m\omega _{c})^{\frac{1}{2}}$ is the
magnetic length, and $E_{0}\!=\!\hbar \omega _{c}/2-g\mu
_{B}B_{z}/2$ is the energy of the ground state ($n\!=\!0$).
In the limit of strong Rashba coupling or large $n$,
$(\alpha/l_{c})\sqrt{2n}\gg E_0$, the spin of the eigenstates lies
in the plane of the 2DEG. These eigenstates have a cyclotron
radius given by
\begin{equation}
r_{c}^{2}\simeq2nl_c^2.
\end{equation}
Then,  states with different $n$, and consequently different
cyclotron radii, coexist within the same energy
window.\cite{UsajB04_focusing} In fact, it is easy to verify that
the difference between the two cyclotron orbits is 
\be
\Delta r_c\simeq \frac{2 \alpha}{\hbar\omega_c}\smeq \frac{\hbar \Delta
k_\pm}{m^* \omega_c}\smeq l_c^2\Delta k_\pm 
\label{deltar} 
\ee
Equivalent results are found in a semiclassical treatment of the
problem.\cite {PletyukhovAMB02,ReynosoUSB04,RokhinsonLGPW04,ZulickeBW07}

Eventually, if both $\alpha$ and $\beta$ were non-zero, one might
hope to be able to gather information on the \textit{shape} of the
Fermi surface by measuring $\Delta r_c$.
As we show below, this is indeed the case when transverse electron
focusing is used to map out $\Delta r_c$ as a function of crystal
orientation or the strength of the Rashba coupling.

\subsection{Transverse electron focusing}

\begin{figure}[t]
 \centering
 \includegraphics[width=.4\textwidth,clip]{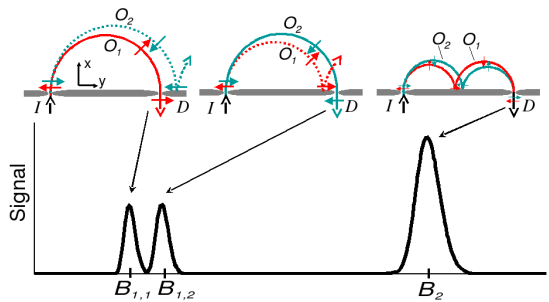}
\caption{Scheme of the focusing experiment in the presence of SO coupling (we take $\beta=0$ for simplicity). An electron injected at QPC $I$ can follow one of two different orbits ($O_1$ and $O_2$) depending on its spin orientations. These two orbits lead to the splitting of the first focusing peak since the corresponding focusing fields $B_{1,1}$ and $B_{1,2}$ are different. After bouncing off the edge the electron that followed orbit the smaller (bigger) orbit continues in the bigger one (smaller one) so in both cases the electron arrives to the detector $D$ for the same focusing field $B_2$, leading to a single peak.}
 \label{focusing}
\end{figure}

The usual geometry for transverse focusing experiments consists of
two quantum point contacts (QPCs) at a distance $L$, which are
coupled to the same edge of a 2DEG (see Fig. \ref{focusing}).
Electrons emitted from QPC $I$ (injector) are focalized onto the
QPC $D$ (detector) by the action of an external magnetic field
perpendicular to the 2DEG. In a classical picture, the electrons
ejected from the injector are forced to follow circular orbits due
to the Lorentz force. If the applied magnetic field has some
arbitrary value, the electrons miss the detector and simply
follow skipping orbits against the edge of the sample. However,
for some particular values of the external field ($B_n$), such
that the distance between the QPCs is $n$ times the diameter of
the cyclotron orbit, with $n$ as an integer number, the electrons reach the detector. 
In such a case, there is a charge
accumulation in the detector that generates a voltage difference
across QPC $D$. This gives voltage peaks as the external field is
swept through the focusing fields
$B_n$.\cite{Tsoi1975,vanHouten89,PotokFMU02}

In a quantum-mechanical description, the scattering states in the
two QPC are coupled by the Landau levels of the 2DEG. As the
Landau eigenstates have a characteristic length (the cyclotron
radius $r_c$) that depends on the applied field, there is also a
matching condition for $B=B_n$.
In this context, the main features of the magnetic-field dependence of the measured signal are contained in the transmission $T$ between the two QPC\cite{BeenakkerH91}---typical experimental setups include also one or two Ohmic contacts at the bulk of the 2DEG which are used to inject currents and measure voltages.\cite{vanHouten89}

As shown in Refs. [\onlinecite{UsajB04_focusing}] and
[\onlinecite{RokhinsonLGPW04}], in systems with either Rashba or
Dresselhaus spin-orbit coupling (but not both) the first focusing
peak splits in two. Such splitting, for $\beta=0$, is given by Eq. (\ref{deltar})
or, in terms of the magnetic field, by $\Delta B\smeq 4 \alpha
m^*c/\hbar e L$, which is independent of $B$. Furthermore,
each peak corresponds to a different spin projection of the
electron leaving the emitter.\cite{UsajB04_focusing} Once again
this can be understood using a classical picture plus the
fact that there are two Fermi surfaces, even though the
semiclassics is not trivial.\cite{LittlejohnF1992,FriskG1993,AmannB2002,PletyukhovAMB02,Zaitsev2002,PletyukhovZ2003,ZulickeBW07}

This simple mechanism, which is able to spatially separate the two spin orientations of a electron beam, was recently used\cite{RokhinsonPW06} to study the current's spin polarization associated with the "$0.7$" anomaly in QPCs and it was also suggested\cite{ReynosoUB07} as a tool to study spin polarization of the flowing current in adiabatic QPCs due to SO.\cite{EtoHK05}

\subsection{Numerical solution}
As mentioned above, we are interested in calculating the
conductance between the two lateral QPCs. In the zero temperature
limit this conductance is just $e^2/h$ times the transmission
coefficient $T$ between the two contacts evaluated at the Fermi
energy. Even in the absence of the spin-orbit interaction, it is
not possible to obtain an analytical solution of the problem when
there is an applied perpendicular magnetic field.
Hence, we calculate $T(E_F)$ numerically using a discretized
system ("tight-binding-like" model) where the leads or contacts
can be easily attached.
 The Hamiltonian of the system can then be written as $H=H_0+H_{SO}$, where
\begin{equation}
H_{0}\! = \!
\sum_{n,\sigma}\varepsilon_{\sigma}^{}c_{n\sigma}^{\dagger}c_{n\sigma
}^{}\smmi\sum_{\langle
n,m\rangle,\sigma}t_{nm}^{}\,c_{n\sigma}^{\dagger}c_{m\sigma}^{}\!
+ \! h.c.\,.
\end{equation}
Here, $c_{n\sigma }^{\dagger}$ creates an electron at site $n$ with spin $%
\sigma$ ($\uparrow$ or $\downarrow$ in the $z$ direction) and energy $%
\varepsilon _{\sigma }^{}\! = \! 4t\! - \! \sigma
g\mu_{B}B_{z}/2$, $t\! = \!  $ $\hbar^{2}/2m^{*}a_{0}^{2}$, and
$a_{0}$ is the effective lattice parameter which is always chosen
to be small compared to the Fermi wavelength. The
summation is made on a square lattice, where the position of the site $n$ is $n_{x}%
\widehat{\bm{x}}+n_{y}\widehat{\bm{y}}$, where $\widehat{\bm{x}}$ and $%
\widehat{\bm{y}}$ are unit vectors in the $x$ and $y$ directions,
respectively. The hopping matrix element $t_{nm}^{}$ is nonzero
only for nearest-neighbor sites and includes the effect of the
diamagnetic coupling through the Peierls
substitution.\cite{Ferrybook} For the choice of the Landau gauge $t_{n(n+%
\widehat{\bm{x}})}\! = \! t\exp{(-\mathrm{i}n_{y}2\pi\phi /\phi _{0})}$ and $%
t_{n(n+\widehat{\bm{y}})}\! = \! t$ with $\phi\! = \! a_{0}^{2}B$
the magnetic flux per plaquete and $\phi _{0}\! = \! hc/e$ the
flux quantum.

The second term of the Hamiltonian describes the spin-orbit coupling,
\begin{eqnarray}
&& H_{SO} \! = \! \sum_{n}\left\{ \lambda_y c_{n\uparrow}^{\dagger}c_{(n+%
\widehat{\bm{y}})\downarrow} - \lambda _y ^* c_{n\downarrow}^{\dagger}c_{(n+%
\widehat{\bm{y}})\uparrow} + \right. \\
&& \left.e^{-\mathrm{i} n_{y}2\pi\phi /\phi_{0}} \left[ \lambda_x
c_{n\uparrow}^{\dagger}c_{(n+\widehat{\bm{x}})\downarrow}\! - \!
\lambda_x
^* c_{n\downarrow}^{\dagger}c_{(n+\widehat{\bm{x}})\uparrow}\right]\right\}%
\! + \! h.c.  \nonumber
\end{eqnarray}
where $\lambda_R \! = \! \alpha /2a_{0}$, $\lambda_D \! = \! \beta
/2a_{0}$, $\lambda_x \! = \! \left(\lambda _{R} \! + \! \mathrm{i}\lambda _{D} e^{-%
\mathrm{i} 2 \varphi} \right)$ and $\lambda_y \! = \! -\left( \lambda _{D} e^{-%
\mathrm{i} 2 \varphi} \!+\! \mathrm{i} \lambda _{R} \right)$ and
$\varphi$ is the angle between the crystallographic axis $x'$ and
the $x$ axis (normal to the edge of the 2DEG). In the second term
the Peierls substitution is made explicit.

Each lateral contact is described by a narrow
stripe with a width of $N_{0}$ sites and, for simplicity, no spin-orbit coupling. They represent point contacts gated to have a single active channel with a conductance $2e^{2}/h$  (for details see Ref. [\onlinecite{UsajB04_focusing}]).
To obtain the conductance between the two contacts we calculate
the retarded (advanced) Green's function matrix,
$\mathcal{G}^{r(a)}$. Because of the lift of the spin degeneracy
the Green's function between two sites $i$ and $j$ has four
components $\mathcal{G}_{i\sigma,j\sigma ^{\prime}}$.

The zero-temperature conductance is then obtained using the Landauer formula,
$
G_{12}\! = \! (e^2/h)\mathrm{Tr}\!\left[\Gamma
^{(1)}\mathcal{G}^{r}\Gamma ^{(2)}\mathcal{G}^{a}\right]
$, evaluated at the Fermi energy. Here $\Gamma ^{(N)}\smeq
\mathrm{i}[\Sigma_{N}^{r}\smmi\Sigma_{N}^{a}]$ is the "coupling
matrix" to the contact $N$ and $\Sigma_{N}^{r(a)}$ the
corresponding self-energies of the retarded (advanced) propagator.


\begin{figure}[t]
 \centering
 \includegraphics[width=.45\textwidth]{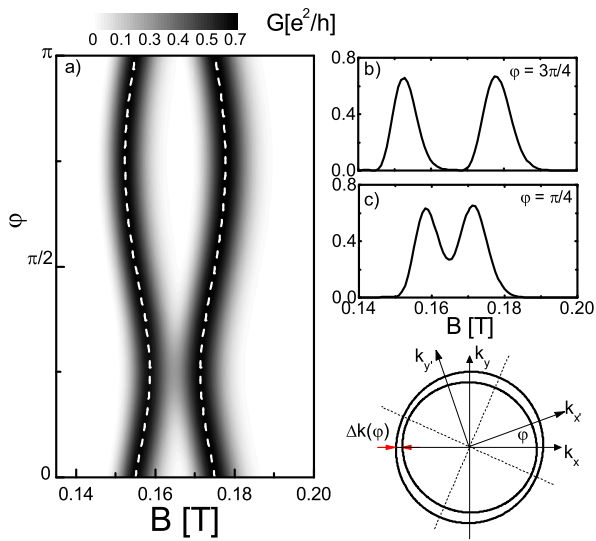}
\caption{(a) Splitting of the first focusing peak as a function the
crystallographic angle $\varphi$ for $E_F=23$meV,
$m^*\smeq0.055m_0$, $L=1.5\mu$m, $\alpha=15$meVnm, and
$\beta/\alpha=1/3$. The dashed lines correspond to Eq.
(\ref{split}) and $B_f\smeq165$mT; (b) and (c) show the focusing
peaks for the minimum and maximum splitting; (d) scheme of the
Fermi surface indicating the difference $\Delta k(\varphi)$.}
 \label{changewithphi}
\end{figure}


\section{Splitting of the focusing peaks}

\subsection{Dependence with the crystal orientation}

Figure \ref{changewithphi} shows the splitting of the first
focusing peak as a function of the crystal orientation (defined by
the angle $\varphi$) with respect to the edge of the sample. The
splitting shows a simple oscillatory behavior, whose angle
dependence can be fully understood in terms of the shape of the
Fermi surface and a simple semiclassical argument presented below. It is worth
mentioning that the semiclassical description of the orbits is far
from trivial in the presence of spin-orbit
coupling.\cite{LittlejohnF1992,FriskG1993,AmannB2002,PletyukhovAMB02,Zaitsev2002,PletyukhovZ2003,ReynosoUSB04,ZulickeBW07} In
particular, in the presence of both Rashba and Dresselhaus
couplings, there is an extra complication due to the possibility to
have mode conversion points (points where the spin-orbit field
cancels).\cite{LittlejohnF1992,FriskG1993,AmannB2002,PletyukhovAMB02,Zaitsev2002,PletyukhovZ2003} This will be important for the effect
discussed in the following section.

Here, however, we can argue that the SO coupling is sufficiently strong so that the spin follows the momentum adiabatically. We can then use the usual semiclassical description for a band structure given by Eq. (\ref{band}). In that case, the semiclassical equations of motion are given by
\begin{equation}
 \dot{\bm{r}}\smeq \bm{v}\smeq \frac{1}{\hbar}\nabla_{\bm{k}}\varepsilon(\bm{k})\,,\qquad \dot{\bm{k}}\smeq \frac{e}{\hbar c} \bm{v}\times \bm{B}\,.
\end{equation}
For $\bm{B}\smeq B\hat{z}$, the solution of these equations implies
that $\bm{k}(t)$ moves along the Fermi surface while
$\bm{r}(t)\smeq\bm{r}(0)\smpl 
\hat{z}\times\left(\bm{k}(t)\smmi\bm{k}(0)\right)l_c^2$ and, as usual,
the real-space orbit is related to the one in $k$-space by a
$\pi/2$ rotation and a scale factor $l_c^2$. Since
 in our case there are two different Fermi surfaces, there are two real-space orbits whose radius
 difference is then given by $\Delta r(\varphi)\smeq l_c^2\Delta k(\varphi)$ where $\Delta k(\varphi)=(2m^*/\hbar^2)\sqrt{\alpha^2+\beta^2-2\alpha\beta\sin{2\varphi}}$ is
the orientation-dependent difference of the two wave vectors of the two
Fermi surfaces (see Fig \ref{changewithphi}).
 Here, in determining the cyclotron radius, we have neglected the fact that the velocity is
 not parallel to $\bm{k}$ (normal injection does not always correspond to $k_y=0$). In the
limit $\Delta k(\varphi)/k_F\ll 1$, as it is the case here, this
is an excellent approximation.

Then, the peak position is given by $B_\pm(\varphi)\smeq B_f(\varphi)\pm \Delta B(\varphi)/2$, where $B_f(\varphi)\simeq \frac{2c}{eL}\sqrt{2Em^*}$ and
\begin{equation}
 \Delta B(\varphi)=\frac{4m^*}{\hbar e L}\alpha\sqrt{1+\left(\frac{\beta}{\alpha}\right)^2-2\left(\frac{\beta}{\alpha}\right)\sin2\varphi}
\label{split}
\end{equation}
is the magnetic-field splitting of the first focusing peak. This
is indicated in Fig. \ref{changewithphi} with dashed lines.

The agreement with the exact numerical result is excellent, showing that the simplified semiclassical picture is very accurate in this regime. From Eq. (\ref{split}), we see that a measure of the peak splitting is a direct way to measure $\alpha$ and $\beta$. One possible way to do it would be to use different sets of pairs of QPCs oriented in different angles with respect to the crystallographic axis.

\begin{figure}[t]
 \centering
 \includegraphics[width=.4\textwidth,clip]{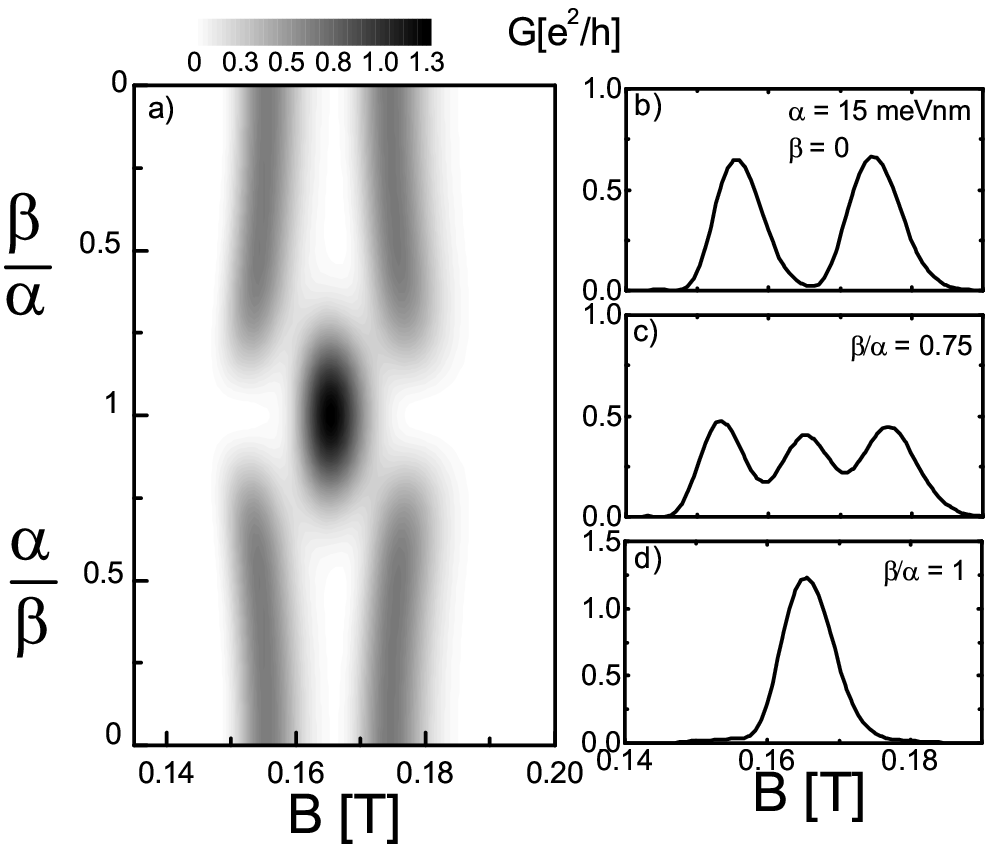}
\includegraphics[width=.2\textwidth,clip]{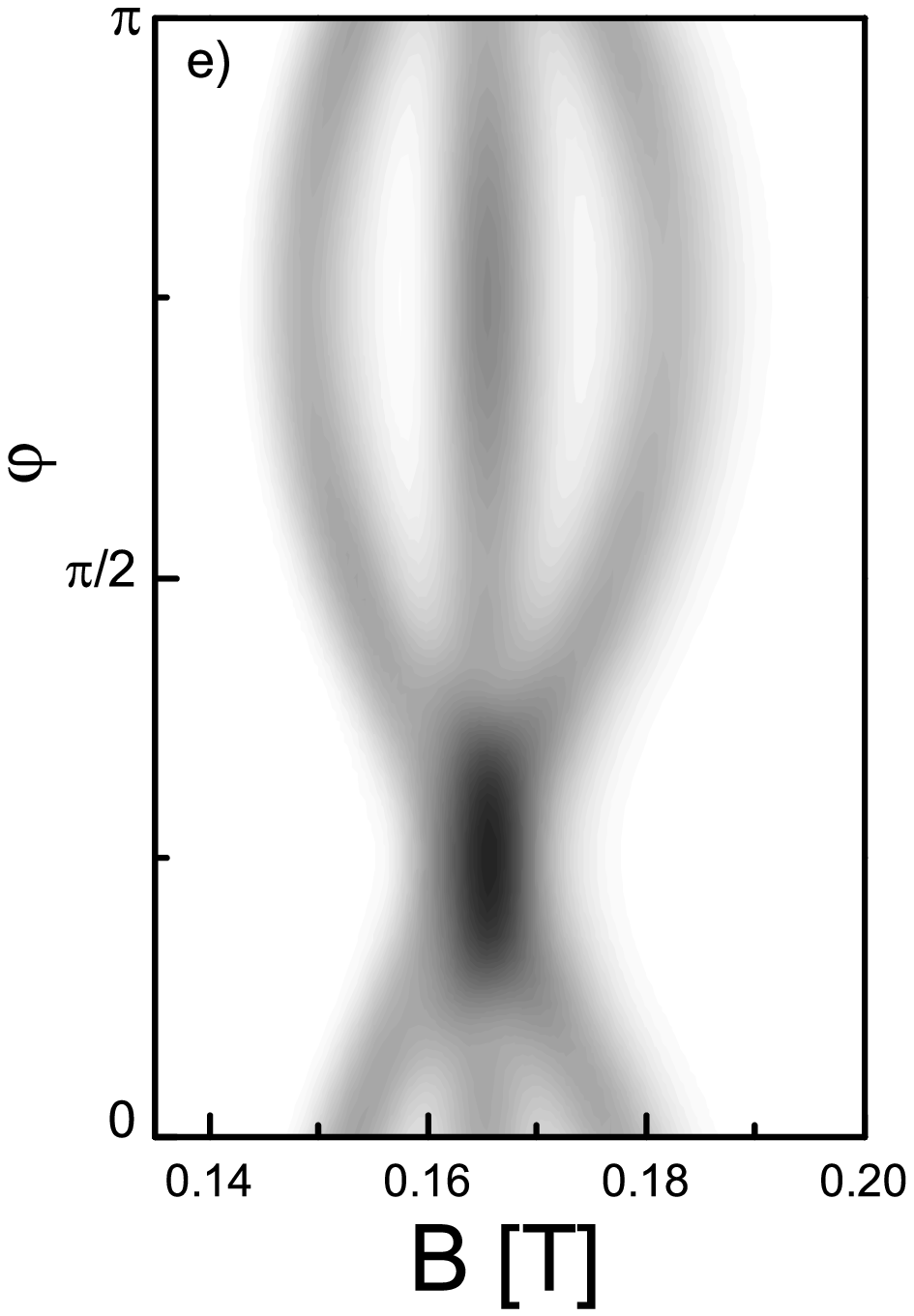}
\includegraphics[width=.2\textwidth,clip]{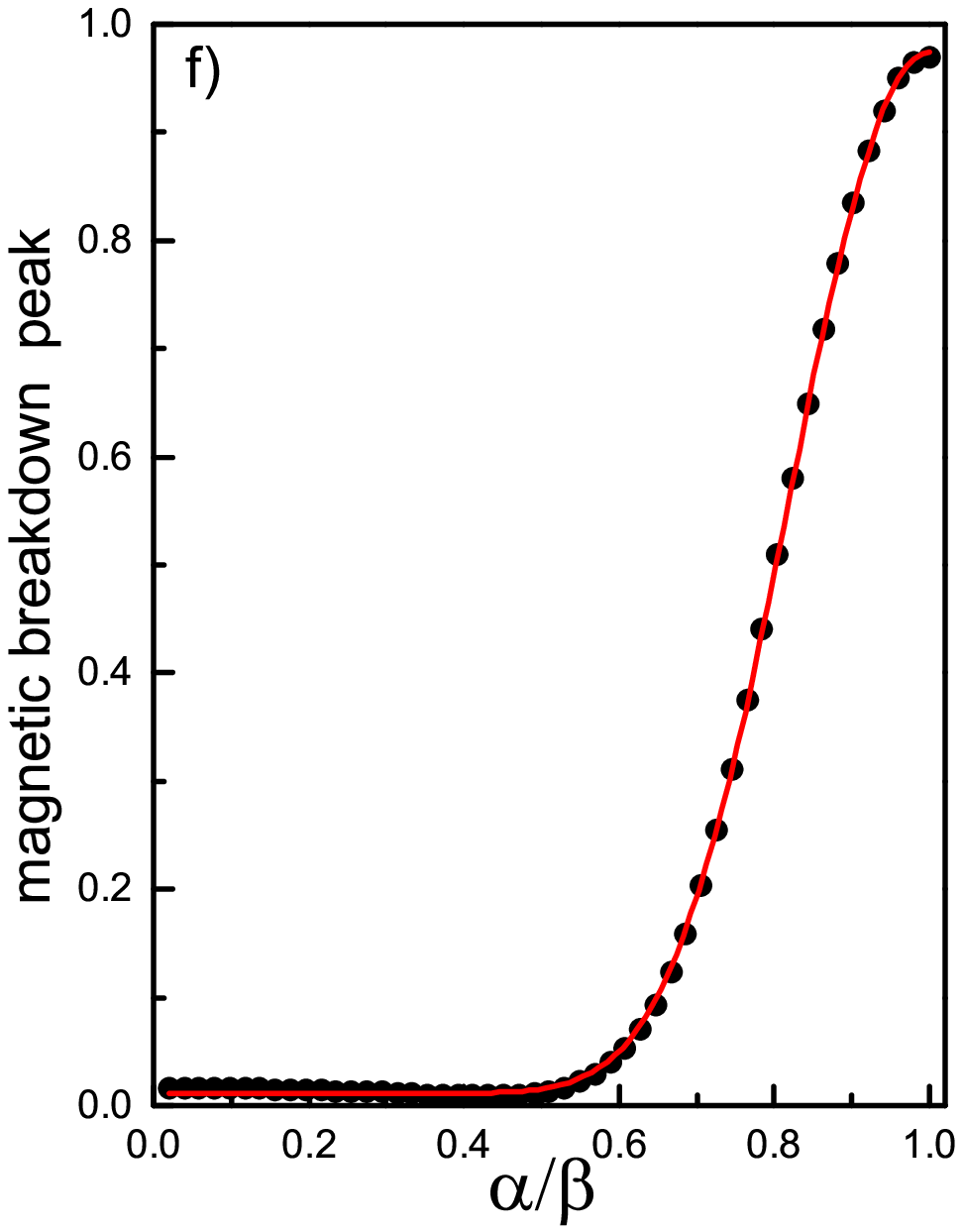}
\caption{Magnetic breakdown between different cyclotron orbits. (a)
Evolution of the first focusing peak as a function of $\alpha$ and
$\beta$ when fixing the greater in $15$ meVnm. The splitting evolves as expected from Eq. (\ref{split})
until the two coupling have similar
values. Close to this point a third peak appears as a consequence of
the quantum tunneling between the two orbits. For $\alpha\smeq
\beta$ we have only one orbit, $\Delta k\smeq0$, and then only one
peak; (b), (c) and (d) show the focusing peaks for different values
of the SO coupling parameters; (e) focusing signal as a function of $\varphi$ for $\alpha/\beta=0.75$; (f) fitting of the amplitude of the
central peak with Eq. (\ref{p}).} \label{magneticbreak}
\end{figure}

\subsection{Additional peak: "magnetic breakdown" }

Let us now consider a different situation where the crystallographic orientation is kept
 fixed (we take $x'\smeq x$ so that  $\varphi\smeq0$) but the magnitude of the spin-orbit
 coupling ($\alpha$ or $\beta$) is changed. This is shown in Fig. \ref{magneticbreak}. When one of the SO couplings dominates
the two peaks structure is clearly seen. Consider the case of
small $\alpha/\beta$ in Fig. \ref{magneticbreak}; as
$\alpha/\beta$ increases the splitting also increases in agreement
with Eq. (\ref{split}). However, for
$\alpha/\beta\simeq 1$, a third peak develops at $B=B_f$. The
amplitude of this peak increases at the expenses of the other two
and becomes the only peak for $\alpha\smeq\beta$. The fact that
there is only one peak when $\alpha\smeq\beta$ is quite clear from
the fact that in that case the two Fermi surfaces are circular and
have the same radius.

The transition from two to three peaks can be understood in terms
of tunneling between cyclotron orbits, in the same spirit as the
magnetic breakdown between band orbits in bulk metals.\cite{CohenF1961} For
$\alpha\simeq\beta$, the "gap" in $k$-space between the two
Fermi surfaces, that is the minimum distance between them $\Delta
k_-\smeq2m^*|\alpha\smmi\beta|/\hbar^2$, is very small and the
magnetic field can induce a tunneling transition between both
orbits. Therefore, if $f(B-B^*)$ describes a single focusing peak
centered around $B^*$, the complete focusing signal is expected to
behave as
\begin{equation}
\left( f(B-B_+)\smpl f(B-B_-)\right)(1-p)+2p\,f(B-B_f)\,,
\label{Landau-Zener}
\end{equation}
where  $p$ is the tunneling probability, which can be estimated
using a Landau-Zener-type argument. Following Refs.
[\onlinecite{Shoenberg1984,AndradaSilvaLRB94}], a rough estimate for $p$ is
\begin{equation}
p=\exp\left(-\pi l_c^2 \sqrt{\frac{\Delta k_-^{3}}{a+b}}\right)\,,
\end{equation}
 where $1/a$ and $1/b$ are the curvature radii of the two Fermi surfaces in the tunneling region, \textit{ie.} close to the minimum gap point. For $\alpha/\beta\sim 1$, this can be approximated by
\begin{equation}
 p=\exp\left(-\frac{\pi m^* k_Fl_c^2 }{\hbar^2} \frac{(\alpha-\beta)^2}{\sqrt{\alpha \beta}}\right)=\exp\left(-\gamma \frac{(x-1)^2}{\sqrt{x}}\right)
\label{p}
\end{equation}
with $x=\alpha/\beta$ and $\gamma=\pi m^* k_Fl_c^2 \beta/\hbar^2$.
This expression fits the numerical data [see Fig.
\ref{magneticbreak}f] up-to a factor of $0.6$ in $\gamma$. Notice
that the fitting function is not a Gaussian and that $p$ is
independent of the crystallographic angle $\varphi$; it only
depends on the local properties of the Fermi surface around the
minimum gap.

At this point it is worth mentioning that the magnetic breakdown
of the cyclotron orbits has previously been used to explain the
anomalous behavior of the magnetoresistance oscillations in
systems with spin-orbit coupling.\cite{AndradaSilvaLRB94} This
interpretation has been challenged very
recently\cite{WinklerPDPS00,KeppelerW2002,Winkler_book} arguing that the dynamics of the
spin cannot follow, in that case, the momentum. Our case, however, is different as we are in the strong spin-orbit limit and the magnetic breakdown interpretation is appropriate.
\begin{figure}[t]
 \centering
 \includegraphics[width=.4\textwidth]{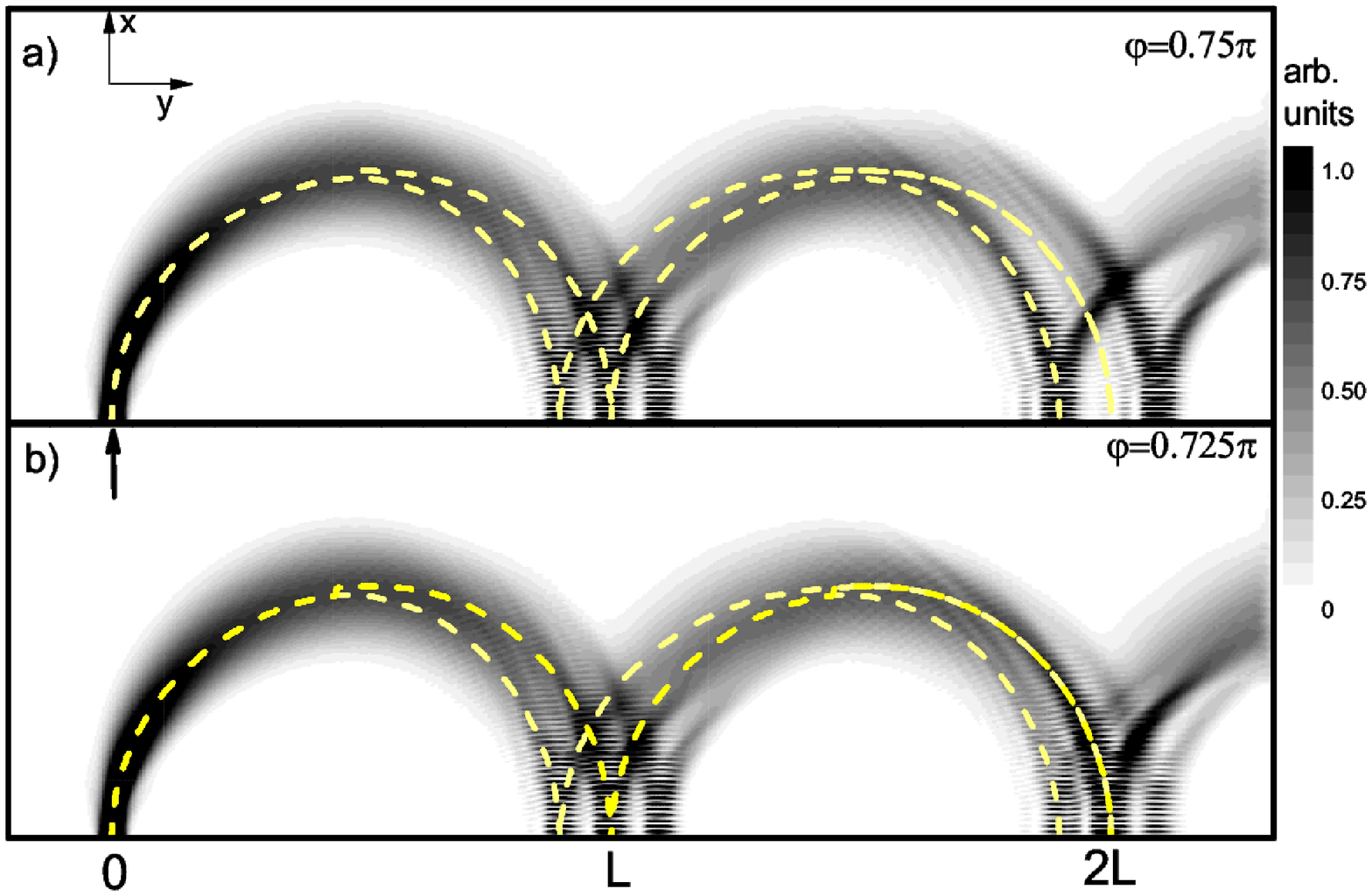}
\includegraphics[width=.4\textwidth,bb=15 14 580 380,clip]{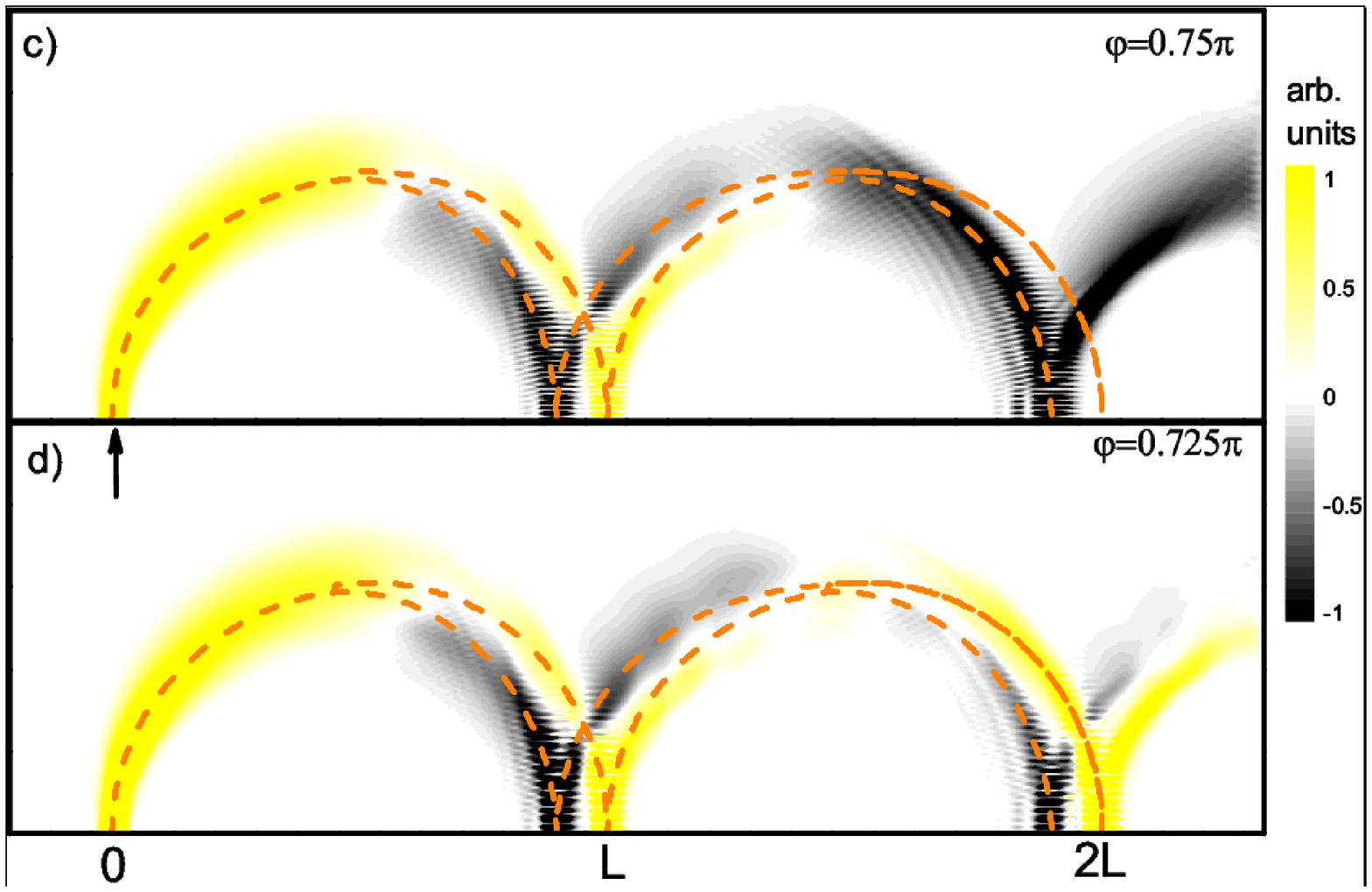}
\caption{Imaging of the cyclotron orbits (see text). (a) and (b)
show the total conductance $G_\mathrm{inj,tip}$ for two different
crystallographic orientation, $\varphi=0.75\pi$ and $0.725\pi$,
respectively, $\beta/\alpha=0.75$ and $L=1.5\mu$m. Note that in both cases
there are three peaks at the first bounce against the edge of the
sample while in the second bounce the central peak is missing (due
to interference) only in (a); (c) and (d) show the spin resolved
conductances,
$G_\mathrm{inj\uparrow,tip\uparrow}-G_\mathrm{inj\uparrow,tip\downarrow}$.
The absence of a sign change along the orbit is a signature of the
magnetic breakdown.} \label{orbits-spatial}
\end{figure}

In order to explicitly show the magnetic breakdown of the cyclotron orbits, we plot in Figs. \ref{orbits-spatial}a and \ref{orbits-spatial}b the conductance $G_\mathrm{inj,tip}$ from the injector to a conducting tip, located above the 2DEG, as a function of the tip position (see Ref. [\onlinecite{UsajB04_focusing}] for details) and for two different crystallographic orientations, $\varphi=0.75\pi$ and $0.725\pi$, respectively. This essentially corresponds to calculate the probability for an electron injected through
 QPC $I$ to reach a given point in the 2DEG and hence brings information about the orbits followed by the electrons.
The images obtained in this way are similar, although with a much better resolution, that the ones we would have obtained by simulating the presence of a tip as a  scatterer (the experimental technique developed in Refs. [\onlinecite{TopinkaLSHWMG00,TopinkaLWSFHMG01,AidalaPHW06,AidalaPKHWHG07}]).

\begin{figure}[t]
 \centering
 \includegraphics[width=.4\textwidth]{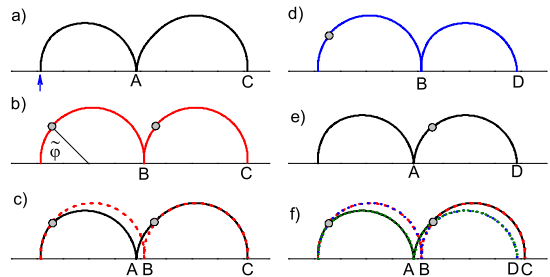}
\caption{Semiclassical orbits (in real space) containing two bounces for an electron injected at the point indicated by the arrow with its spin polarized along the edge of sample. (a) Direct orbit (no tunneling) and (b) orbit with two tunneling events, indicated by the circles. Here $\tilde{\varphi}$ defines the angle where tunneling occurs; (c) superposition of orbits (a) and (b). Notice that after the second tunneling event the two orbits overlap; (d) and (e) orbits that involve one tunneling event; (f) superposition of all previous orbits. The letters A,B,C and D identify the different locations of the bounces.} 
\label{schemeorbits}
\end{figure}

Although it is difficult to distinguish the orbits along the full path, three different orbits are apparent in both cases close to the first bouncing point, which correspond to the first focusing condition. 
As each of these orbits have a spin projection associated with it, we plot in Figs. \ref{orbits-spatial}c and \ref{orbits-spatial}d the difference between the spin resolved conductances, $G_\mathrm{inj\uparrow,tip\uparrow}-G_\mathrm{inj\uparrow,tip\downarrow}$, where the spin-quantization axis corresponds to $\hat{\bm{y}}$. This allows us to follow the direct orbits and the ones that involve tunneling. For this, it is important to take into account that for $\beta=0$, we would observe a change of sign of $G_\mathrm{inj\uparrow,tip\uparrow}-G_\mathrm{inj\uparrow,tip\downarrow}$ in the middle of the orbit as the spin rotates from $|\uparrow\rangle$ to $|\downarrow\rangle$. Here, the fact that there is an orbit in which this quantity does not change sign is an indication of the tunneling from one orbit to the other.
To support this interpretation, we also show the semiclassical orbits (dashed lines) expected for an electron injected with a given spin polarization along the $y$ axis. To include the orbits that involve tunneling, we simply change from one orbit to the other at the position related to the minimum gap in $k$-space (see Fig. \ref{schemeorbits}).


\begin{figure}[t]
 \centering
 \includegraphics[width=.4\textwidth]{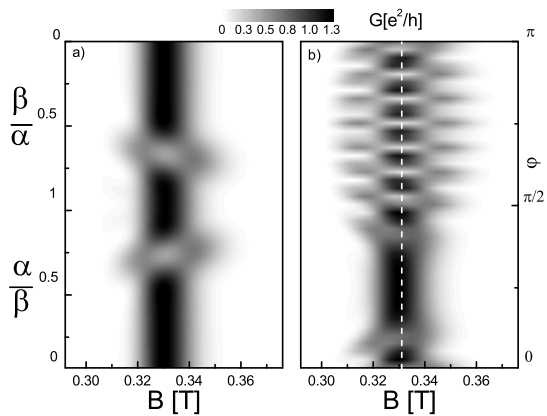}
\caption{Focusing signal at the second focusing peak: (a) as a
function of $\alpha$ ($\beta$) for $\beta(\alpha)=15$meVnm and
$\varphi=0$. As expected there is only a single peak, except for
$\alpha/\beta\simeq 0.75$ where there is a destructive
interference between the different orbits shown in Fig. \ref{schemeorbits}. Notice this
point correspond to the case where there are three peaks at the
first focusing condition; (b) as a function of the crystallographic
angle $\varphi$ for $\alpha/\beta\simeq 0.75$.}
\label{secondpeak}
\end{figure}


An interesting effect occurs at the second focusing peak (second
bounce in Fig. \ref{orbits-spatial}). In such a case, the peak
structure results from the interference of several paths and hence
destructive interference can results for particular values of the
parameters. In particular, we see in Fig. \ref{orbits-spatial} that for $\varphi=0.75\pi$ the central peak is missing. 
In order to understanding the origin of this effect, we show in Fig. \ref{secondpeak}a the focusing signal as a function of $\alpha/\beta$ for $\varphi=0$ and the same microscopic
parameters as in Fig. \ref{magneticbreak}.
For almost all values of $\alpha/\beta$ there is a single peak (as
in the $\beta=0$ case). However, for $\alpha/\beta\simeq 0.75$,
when the first focusing peak shows a three peak structure (see Fig. \ref{magneticbreak}), the
central peak disappears while two satellites peaks emerge. When
setting $\alpha/\beta=0.75$ this effect has an oscillatory
behavior as a function of the crystal orientation. This is shown in Fig.
\ref{secondpeak}b (see also Fig. \ref{magneticbreak}(e) for a comparison with the first focusing peak structure). 

As mentioned above, the origin of this modulation of the amplitude of the second focusing peak is the interference between the different orbits that contribute to the signal.
These orbits, for an electron injected with its spin pointing along the $y$-axis, are shown in Fig. \ref{schemeorbits}. The regions in real space where tunneling
between the two orbits occurs are indicated with circles. The
position of these regions depends on the crystal orientation. Figures \ref{schemeorbits}a and \ref{schemeorbits}b show the orbits that contribute to the central peak of the second focusing peak, while Fig. \ref{schemeorbits}c shows the overlap of the two orbits.
Figures \ref{schemeorbits}d and \ref{schemeorbits}e show the orbits that contribute to one of the satellites in the second focusing peak.  The letters A,B,C and D identify the different locations of the bounces and then the different focusing peaks.
%

As the phase acquired due to the tunneling between orbits is
independent of $\varphi$, in order to account for the angle
dependence of the interference pattern we need to include  the
orbital phase acquired by the electron along the different paths. 
In addition, we notice that after the second tunneling event the two orbits in Fig. \ref{schemeorbits}c follow exactly the same path. Since they accumulate the same orbital phase from there to the next bouncing point, if they interfere destructively right after the tunneling, they will do it along the final part of the orbit. 
This is
clearly seen in Fig. \ref{orbits-spatial} where for
$\varphi=0.75\pi$ the orbit corresponding to the central peak in
the second focusing peak has disappeared completely after the
second tunneling event.

The central peak amplitude is then given by
\begin{equation}
 A=|\sqrt{P_d}+\sqrt{P_t}e^{\ci(2\theta_t+\theta_{orb})}|^2
\label{A}
\end{equation}
where $P_d$ and $P_t$ are the probability of the direct and
tunneling paths, respectively [paths (a) and (b) in Fig. \ref{schemeorbits}], $\theta_t$ is the phase acquired by the electron due to each
tunneling event (we assume them to be equal), and $\theta_{orb}=\theta_1-\theta_2$ is the difference of the orbital phases of the two paths.
In the semiclassical picture, the orbital phase of each path is given by
$1/\hbar \int_C \bm{p}\cdot \mathrm{d}\bm{r}$, where the integral
is done along the classical path $C$, $\bm{p}$ is the canonical
momentum and $\bm{r}$ the position vector satisfying
\begin{equation}
\dot{\bm{r}}=\frac{\partial H}{\partial \bm{p}}\qquad
\dot{\bm{p}}=-\frac{\partial H}{\partial \bm{r}}
\end{equation}
In the strong spin-orbit limit, the Hamiltonian $H$ is given by\cite{AmannB2002,PletyukhovZ2003,ZulickeBW07}
\be
H=\frac{P^2}{2m^*}\pm \sqrt{(\alpha P_{x'}\smpl\beta P_{y'})^2\smpl(\alpha P_{y'}\smpl\beta P_{x'})^2}
\ee
with $\bm{P}=\bm{p}+(e/c)\bm{A}$.  
Using a symmetric gauge, it is straightforward to show that
$1/\hbar \int_C \bm{p}\cdot \mathrm{d}\bm{r}=(1/2l_c^2)\int
r_C^2 \mathrm{d}\phi$, where $r_C$ is the radius of the
corresponding cyclotron orbit. The orbital phase difference between the two paths is then
\begin{equation}
\theta_{orb}=\frac{1}{2l_c^2}\left(\int_0^{\tilde{\varphi}}
(r_1^2-r_2^2)\,\mathrm{d}\phi-\int_{\tilde{\varphi}}^\pi
(r_1^2-r_2^2)\,\mathrm{d}\phi\right)
\label{theta-exact}
\end{equation}
with $\tilde{\varphi}$ as the angle where tunneling occurs (see Fig. \ref{schemeorbits}) and $r_i$, $i=1,2$, as
the radii of the orbits. This integral can be calculated
analytically for arbitrary values of $\alpha$ and $\beta$ in terms
of elliptic integrals. However, for $\alpha/\beta\simeq1$, it can
be approximated by
\begin{equation}
\theta_{orb}=4l_c^2k_F\frac{2\sqrt{\alpha\beta}m^*}{\hbar^2}\cos(\varphi-\frac{\pi}{4}).
\label{phase}
\end{equation}
\begin{figure}[t]
 \centering
 \includegraphics[width=.4\textwidth]{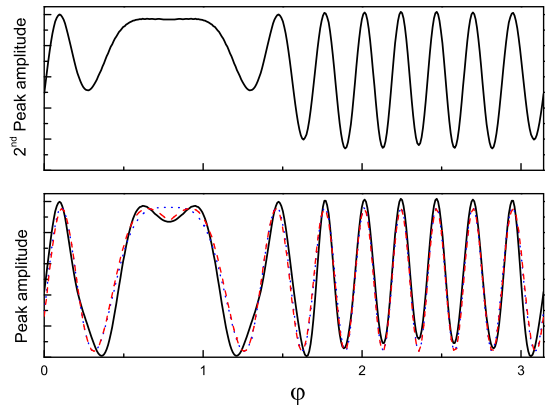}
\caption{(a) Amplitude of the second focusing peak (indicated by a
dashed line in Fig. \ref{secondpeak}) as a function of the
crystallographic orientation. The oscillation is due to quantum
interference between different paths (see Fig. \ref{schemeorbits}); (b) same as above but without
taking into account the contribution from the satellite peaks. The
dashed (dotted) line corresponds to a fitting using the exact
(approximated) value of $\theta_{orb}$, given by Eq. (\ref{theta-exact})[Eq. (\ref{phase})].} \label{interference}
\end{figure}
Figure \ref{interference} shows the total central peak amplitude
obtained numerically as a function of $\varphi$. Since for
$\varphi\simeq\pi/4$ the satellite peaks merge into the central
peak, in Fig. \ref{interference}b we plot the central peak
amplitude subtracting the satellite contributions. This can be
done because the orbits that contribute to the central peak for
all angles are those where the electron arrives with the same spin
orientation it had at the injector. A fitting of the peak
amplitude, using Eq. (\ref{A}) with $P_d$ and $P_t$ as the only
fitting parameters, is shown in Fig. \ref{interference}b. The dotted line was obtained using the approximated expression for $\theta_{orb}$ [Eq. (\ref{phase})], while the dashed line corresponds to the exact expression [Eq. (\ref{theta-exact})]. Once again, the
agreement is very good, given further support to the magnetic
breakdown interpretation.

It is easy to verify that for the case of the satellite peaks,
$\theta_{orb}$ takes the same value than in the case of the
central peak. However, as in this case both interfering paths contain a
tunneling event, $\theta_t$ drops out. Then, the fact that there
is destructive interference in the central peak when the satellites
have there maximum values indicates that $\theta_t\simeq\pi/2$ as
expected from the magnetic breakdown picture.
In the above analysis, we did not consider explicitly the phase acquired by the spin degree of freedom. In general, such phase has a nontrivial dependence with the geometry of the orbit. \cite{LittlejohnF1992,FriskG1993,AmannB2002,PletyukhovAMB02,Zaitsev2002,PletyukhovZ2003} Here, since we assumed that the spin follows the SO field, that is, it rotates around the $z$ axis, this phase is not relevant. However, special care should be taken when considering the effect of the tunneling between orbits as the spin may have an additional rotation. As this case involves a mode conversion point, a semiclassical analysis of this phase is not simple--- we are not aware of a simple way to estimate it. Therefore, the fact that our  results indicate that the phase due to tunneling is just the usual $\pi/2$, is an indication that the spin phase is $0$ or $\pi$. Further work is needed to clarify this point.

\section{Summary}

We have shown that the interplay between the Rashba and Dresselhaus couplings introduces new effects on the transverse electron focusing. The most interesting aspect is the appearance of additional structure of the focusing peaks related to the magnetic breakdown of the cyclotron orbits when the two SO couplings have similar magnitude: the two sheets of the Fermi surface lead to different paths in real space and, as we have shown,  the tunneling between different paths generates new structure in the focusing peaks. In addition, interference effects between these paths lead to an oscillatory behavior of the second focusing peak amplitude as a function of the orientation of the crystallographic axes. We have shown that the observed interference effects are dominated by the orbital phase accumulated along the different paths. This is so because in this regime, the spin adiabatically follows the momentum and the associated Berry phase cancels out while the spin phase due to tunneling seems to be irrelevant here. One could however envision a different regime where the spin dynamics would be important for which a deeper understanding of the spin dynamics during the tunneling as well as of the semiclassical description of the problem is needed.

The magnetic breakdown of the orbits as well as the interference effect could be directly observed using some of the techniques recently developed for imaging the electron flow. \cite{TopinkaLSHWMG00,TopinkaLWSFHMG01,AidalaPHW06,Reynoso2006b}
In this work we have been mostly concerned with the dependence of the different effects on the crystal orientation. In practice, the experimental observation is a bit cumbersome as it requires us to tailor different QPC setups in different orientations. An alternative to this could be the use of in-plane magnetic fields to modulate $\Delta k_-$ as well as $k_+$ and $k_-$. For instance, for $\alpha\simeq\beta$, an in-plane field applied along the $\hat{x'}-\hat{y'}$ direction  will essentially control the magnitude of the gap in $k$-space, and then the probability for tunneling. On the other hand, in-plane field along the $\hat{x'}+\hat{y'}$ direction can be used to modulate $k_+$ and $k_-$ and then the interference pattern. This also have the advantage of keeping  $k_+ +k_-$ constant and then the focusing condition.

Finally, it is  interesting to note that while the splitting of the focusing peaks allows for the spatial separation of the spin components as in a Stern-Gerlach device,\cite{UsajB04_focusing,RokhinsonLGPW04} the presence of the magnetic breakdown between orbits provides a way to separate a given spin component in a superposition of two spatially separated orbits.

\section{Acknowledgments}

This work was supported by ANPCyT Grants N$^o$13829, 13476, 2006-483
and CONICET PIP 5254. A.A.R. and G.U. acknowledge support from CONICET.


\end{document}